\documentclass[aps,prl,reprint,superscriptaddress]{revtex4-1}

\usepackage[T1]{fontenc}
\usepackage{longtable}
\usepackage{morefloats}
\usepackage{color}
\usepackage{graphicx,amsfonts,amsbsy}
\usepackage{amsmath,amsfonts,amsthm,amssymb}
\usepackage{appendix}
\usepackage{makeidx}
\usepackage{url}
\usepackage{verbatim}
\usepackage[bookmarksnumbered,pdfpagelabels=true,plainpages=false,colorlinks=true,linkcolor=blue,citecolor=blue,urlcolor=blue]{hyperref}
\usepackage[rightcaption]{sidecap}
\usepackage{array}
\usepackage{booktabs}
\usepackage{multirow}
\usepackage{tabularx}
\usepackage{cancel,soul,ulem}

\usepackage{mathrsfs}

\renewcommand{\vec}{\mathbf}
\newcommand{\kp}{{\vec k_\parallel}}

\begin{document}
\title{Even-odd effect for spin current through thin antiferromagnetic insulator}

\author{Niklas Rohling}
\affiliation{Department of Physics, University of Konstanz, D-78457 Konstanz, Germany}

\author{Roberto E. Troncoso}
\affiliation{Center for Quantum Spintronics, Department of Physics, Norwegian University of Science and Technology, NO-7491 Trondheim, Norway}
\affiliation{School of Engineering and Sciences, Universidad Adolfo Ib\'a\~nez, Santiago, Chile}

\begin{abstract}
Magnon spin transport in a metal-antiferromagnetic insulator-ferromagnetic insulator heterostructure is considered. The spin current is generated  via the spin Seebeck effect and in the limit of clean sample where the effects of interface imperfections and lattice defects are excluded. For NiO as an antiferromagnetic insulator we have a magnetic order of antiferromagnetically combined planes {which are internally in ferromagnetic order}. We find that the sign of the spin current depends on the magnetization direction of the plane next to the metal resulting in an even-odd effect for the spin current. Moreover, as long as damping is excluded, this even-odd effect is the only remaining dependence on the NiO thickness for high temperatures.
\end{abstract}

\maketitle
 
\begin{figure}
\centering
\includegraphics{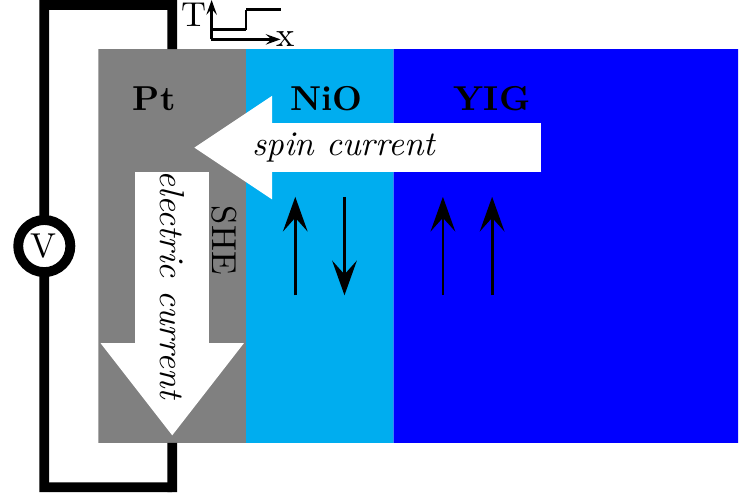}
    \caption{Schematic setup consisting of a NM-AF-F trilayer heterostructure. As a model system it is considered a platinum (Pt), a thin antiferromagnetic NiO layer, and yttrium iron garnet (YIG) film, which is in good approximation a ferromagnet. A thermal gradient yields spin transport through the metal-insulator interface, which causes an electric current in Pt due to the inverse spin Hall effect (SHE) which allows detection of the spin current by electrical means.}
    \label{fig:trilayer}
\end{figure}

\textit{Introduction.}--- Leveraging spin-angular momentum in magnetic insulators gathers considerable interest due to intrinsic low-dissipative transport properties. Spin transport experiments on heavy metal-ferromagnet (HM-F) heterostructures have shown an enhancement when a thin NiO antiferromagnetic layer is placed in between, {forming}, e.g., a platinum{-NiO-} yttrium iron garnet (YIG) {tri-}layer \cite{Wangetal2014,Wangetal2015,Linetal2016}, see Fig.~\ref{fig:trilayer}. Such an enhancement is described by theoretical models both in the diffusive limit \cite{Rezende_et_al2016} and when transport is governed by evanescent spin waves in the NiO layer \cite{Khymyn_et_al2016}. However, various issues are far from being understood, such as the relation between the crystal and the magnetic configurations of the AF, as well as interfacial properties at each contact. Note that there was a significant sample dependence in the experiments which might be due to the properties just mentioned varying from sample to sample. Further experimental investigations related to spin transport through NiO layers include the study of spin Hall magnetoresistance in ferromagnetic insulator-NiO-heavy metal sytems \cite{Shangetal2016,Dongetal2019,Zhuetal2022}, spin transport from a ferromagnetic to a nonmagnetic metal \cite{Zhuetal2021} and from one magnetic metal to another \cite{Dabrowskietal2020}, as well as non-local spin transport in a Pt-NiO-YIG trilayer \cite{Hoogeboometal2021}.

The spin current in Refs.~\cite{Wangetal2014,Wangetal2015} were generated via ferromagnetic resonance by a microwave field. Using a thermal gradient to produces spin currents via the spin Seebeck effect instead, as demonstrated in Ref.~\cite{Linetal2016}, allows further insights as more magnon modes are involved in the transport of spin. Further, spin-Seebeck-effect transport experiments have been performed with NiO \cite{Prakashetal2016} and a metallic AF in between the HM-F structure \cite{Crameretal2018}. In contrast to \cite{Linetal2016}, the reports at \cite{Prakashetal2016,Crameretal2018} report the absence of enhancement of spin currents compare to the HM-F bilayer for most of the parameters. What was found in all three works \cite{Linetal2016,Prakashetal2016,Crameretal2018} was a thickness-dependence of the peak temperature, which { is the temperature} allowing the strongest spin transport. The peak temperature was found to be increasing with thickness of the antiferromagnetic layers. In contrast, in Ref.~\cite{Baldratietal2018} studying epitaxial NiO in $[001]$ direction, no peak temperature was found up to room temperature, which is attributed to a higher N\'eel temperature for epitaxial NiO. Furthermore, Ref.~\cite{Baldratietal2018} reports no spin transport enhancement by the NiO layer for YIG-NiO-Pt, but for Fe$_3$O$_4$-NiO-Pt trilayers.

In this letter, we investigate how the AF layers oriented in (111) direction -- namely the number of atomic planes in transport direction -- impact the spin current propagation. We focus on the spin Seebeck effect across a clean and ideal interface. We find that the sign of the spin current is determined by the number of atomic NiO planes being even or odd. Additionally, we find that in the limit of high temperatures, this even-odd effect is the only remaining dependence on the NiO thickness in the clean limit considered here. This is due to the normalization condition for magnons as bosonic modes.

\textit{Model.}-- We consider a trilayer system with the antiferromagnetic layers(NiO) in between, with the $x$ axis parallel to the (111) direction, orthorgonal to interface and the hard in-plane axis along $y$ \cite{Haakonetal}. The spin Hamiltonian of the system is,
\begin{align}
H_{\rm AF} =\nonumber \frac{1}{2}\sum_{\vec r\in {\rm AF}}&\left[{J_1}\sum_{\boldsymbol\delta} \vec S_{\vec r}\cdot\vec S_{\vec r+\boldsymbol\delta}+{J_2}\sum_{\boldsymbol\eta} \vec S_{\vec r}\cdot\vec S_{\vec r+\boldsymbol\eta}\right.\\
&\left. +  2D_1(S^x_{\vec r})^2 + 2D_2(S^y_{\vec r})^2\right],
\end{align}
where the exchange interaction  $J_1<0$ and $J_2>0$ are the exchange couplings to the nearest and next-nearest neighbors, respectively. The biaxial magnetocrystalline anisotropy is parametrized by the strengths $D_1$ and $D_2$ and the vectors joining nearest and next-nearest neighbors are $\boldsymbol\delta$ and $\boldsymbol\eta$, respectively. The magnetic parameters we use, are
$J_1 = -16$ K, $J_2 = 221$ K, $D_1 = 1.13$ K, and $D_2 = 0.06$ K \cite{HutchingsSamuelsen1972}. The spins system in the ferromagnetic insulator share the same fcc lattice as the AF to avoid the influence of interface roughness and lattice mismatch. Thus the Hamiltonian is,
\begin{equation}
    H_{\rm FI} = \frac{1}{2}\sum_{\vec r\in \rm FI}\left[{J_F}\sum_{\boldsymbol\delta}\vec S_{\vec r}\cdot\vec S_{\vec r+\boldsymbol\delta} -2K_F(S^z_{\vec r})^2\right],
\end{equation}
with $J_F$ the exchange coupling and $K_F$ the uniaxial easy-axis anisotropy. The parameters are choosen such that the saturation magnetization and low-energy dispersion of YIG is matched, with a (non-physical) spin quantum number of $S_F=0.16$ and $J_F=-400\,$K, as well as a small anisotropy $K_F=10^{-5}|J_F|$. The AF-FI interaction is
\begin{equation}
    H_{\rm AF-FI} =  J_{I1}\sum_{\stackrel{\vec r\in{\rm AF}}{\vec r+\boldsymbol\delta\in {\rm FI}}}\vec S_{\vec r}\cdot\vec S_{\vec r+\boldsymbol\delta} + J_{I2}\sum_{\stackrel{\vec r\in{\rm AF}}{\vec r+\boldsymbol\eta\in{\rm FI}}}\vec S_{\vec r}\cdot\vec S_{\vec r+\boldsymbol\eta},
\end{equation}
with $J_{I1,I2}$ the interfacial antiferromagnetic exchange coupling. The metal is described by a simple tight-binding model, $H_M=-{t}\sum_{\sigma=\uparrow,\downarrow}\sum_{\vec r,\boldsymbol\delta}\left(c^\dag_{\vec r\sigma}c_{\vec r+\boldsymbol\delta,\sigma} + h.c.\right)/2$, where $\boldsymbol\delta$ is again the vector to a nearest neighbor and the hopping energy is assumed to be  $t=1\,$eV$=1.16\times10^4\,$K.

\begin{figure}
\includegraphics[width=\columnwidth]{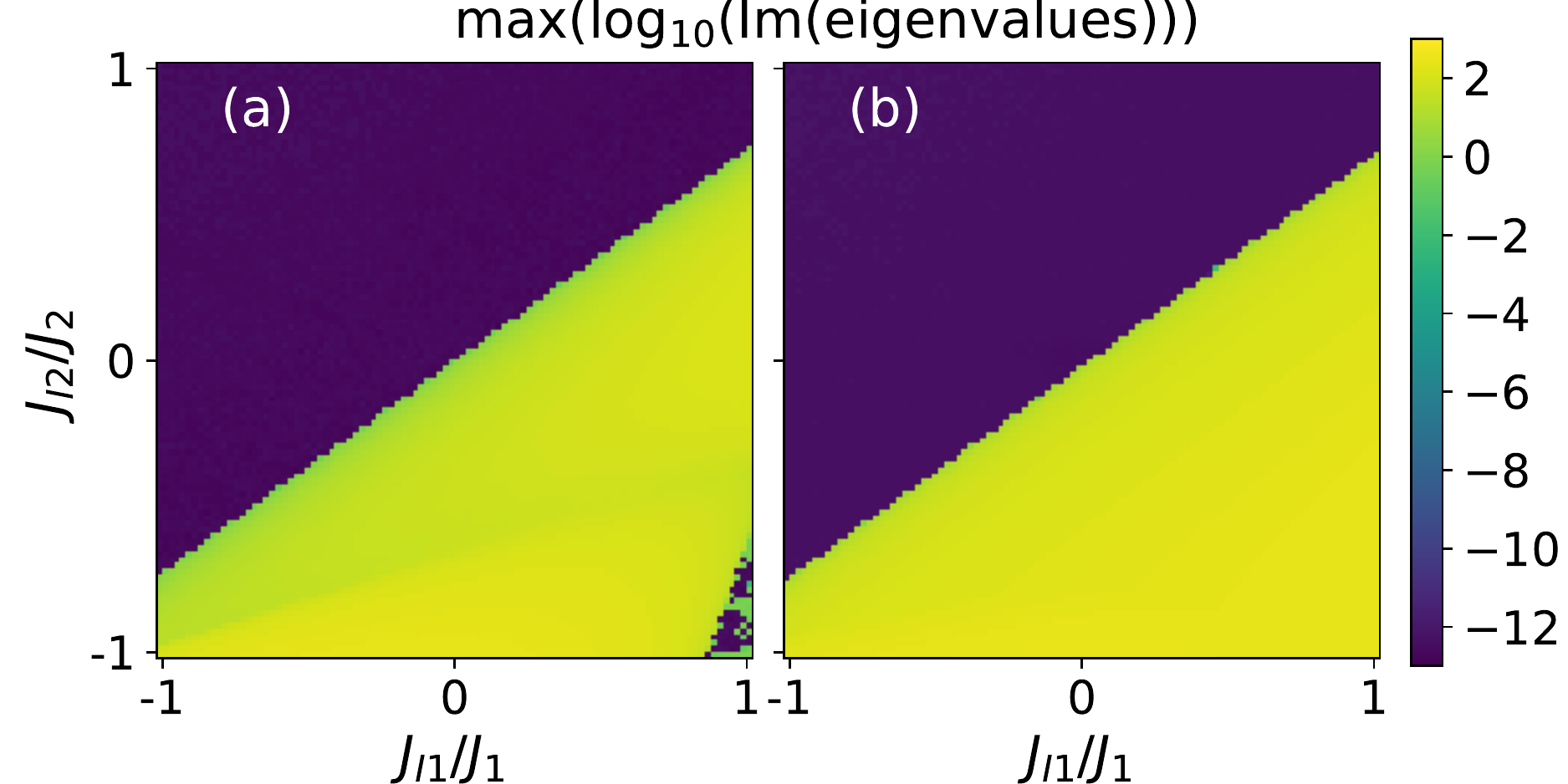}
\caption{(a) Stability of the magnetic configuration where the spins of a single NiO layer are pointing in $-z$ direction while in the five atomic FI layers the spins are pointing in $+z$ direction. Imaginary components of the eigenvalues of the dynamical matrix, see Suppl.~Mat.~for details, around $10^{-12}$ originate in the numerical calculations and indicate regions of potential stability for this configuration. (b) Same as in (a) but now with four NiO layers having their spins pointing in $+z$ and $-z$ direction, respectively.}
\label{fig:StabilitySinglelayer}
\end{figure}

\textit{Magnetic stability and dynamics.}-- We now investigate the stability of the magnetic configuration of the joint NiO-ferromagnet system. We consider the AF and the FI as a joint system, $H_I=H_{\rm AF} +H_{\rm FI} + H_{\rm AF-FI}$, and determine the classical ground state, shown at Fig. \ref{fig:StabilitySinglelayer}, where the spins point in $+z$ or $-z$ direction, while the configuration with spins pointing out of the interface plane are energetically not desired. Note that we have a different number of layers with B ($-z$ spin direction) and A ($+z$ spin direction) configuration as the spins in the FI belong all to A configuration. Note further that in contrast to the typical situation of an antiferromagnet, our system is also not invariant under interchanging the A and the B sublattices within the NiO layer (even if the number of layers in $x$ direction us even) because of the coupling to the FI.

Later, we study the low-energy spin dynamics around the classical ground state. We perform a Holstein-Primakoff transformation \cite{HP} around the classical ground state and truncate after second order in creation and annihilation operators. Then, we continue with a Fourier transform in $yz$ plane. Finally a multi-flavour Bogoliubov transformation, see e.g. \cite{Smitetal2020}, yields the magnonic eigenstates of the system. We solve the remaining eigenvalue problem for any in plane magnon momentum, $\vec q_\parallel = q_y\vec e_y + q_z+\vec e_z$, numerically and label the eigenstates formally by $q_x$. This also allows to investigate the stability of the initially guessed ground state, as imaginary components in the eigenenergy indicate that this state was unstable, see Suppl.~Mat.~for details.

\begin{figure}
    \centering
    \includegraphics[width=\columnwidth]{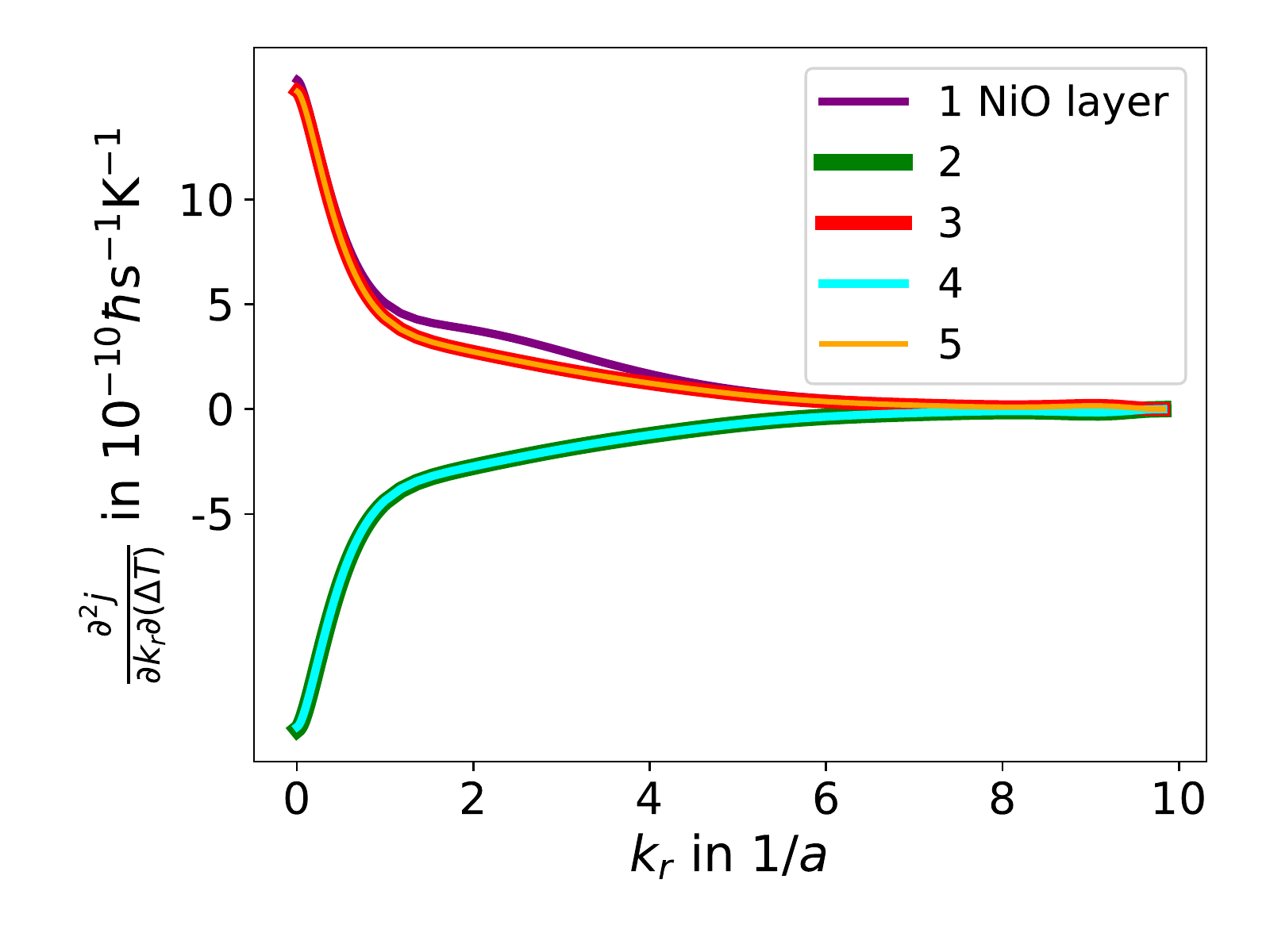}
    \caption{Differential spin current density per thermal gradient, $\partial^2j/(\partial k_r\partial(\Delta T))$, as a function of the in-plane absolute momentum transfer $k_r=\sqrt{(\Delta k_y)^2 + (\Delta k_z)^2}$ at $T=300\,$K for different number of atomic layers of NiO.
    The number of ferromagnetic layers is always five.
    Note the qualitative difference between even and odd.}
    \label{fig:diffcurrent}
\end{figure}

\begin{figure}
    \centering
    \includegraphics[width=\columnwidth]{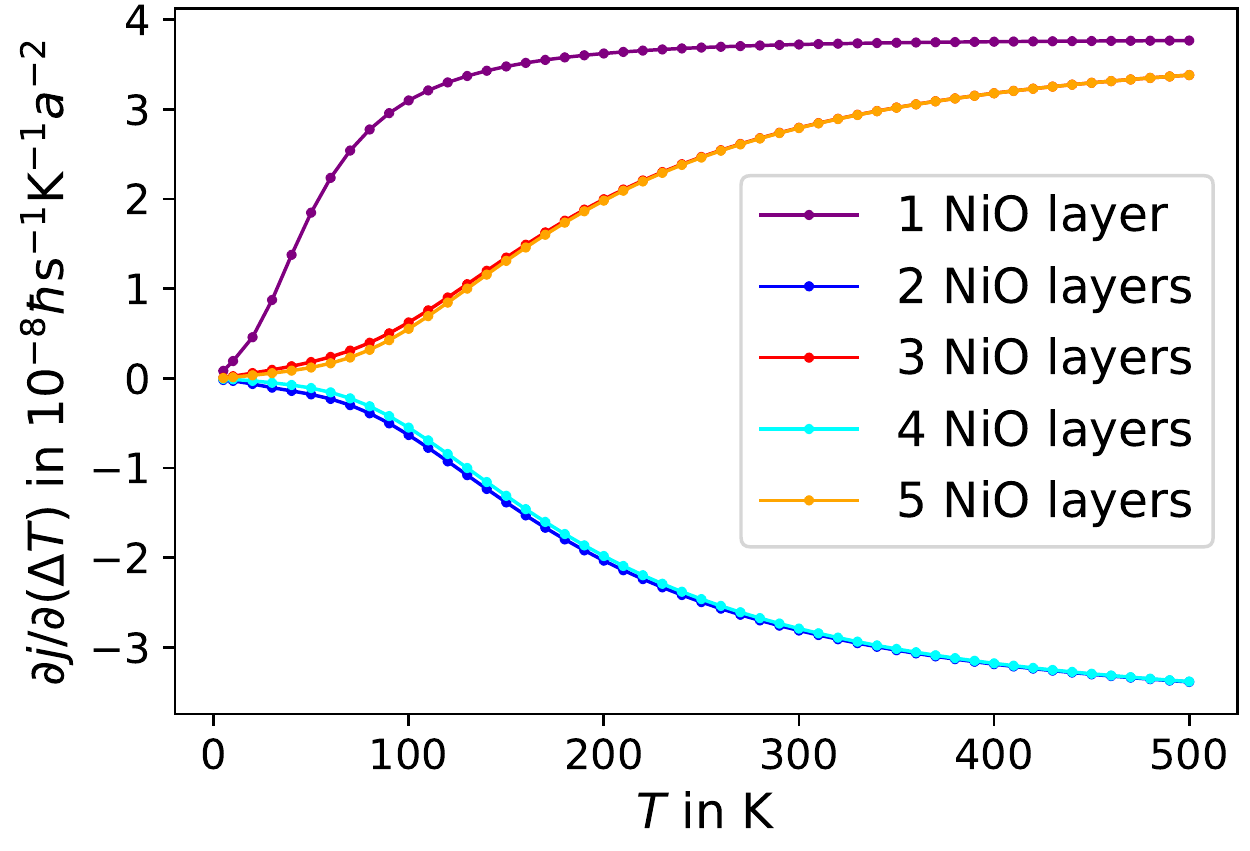}
    \caption{Spin current density per thermal gradient, i.e., the Spin-Seebeck coefficient, for an exchange coupling to the metal of $J_I=1\,$K, $a$ is the lattice constant of NiO.
    We see that for increasing temperatures, the only remaining thickness dependence is the even-odd effect, i.e., the sign of the spin current density depending on the number of NiO atomic planes being even or odd.}
    \label{fig:tempdep}
\end{figure}

\textit{Spin Seebeck effect.}-- We now compute the magnon spin current based on the Fermi's Golden Rule formalism \cite{Benderetal2012,Fjaerbuetal2017}. The transition from an initial ($i$) and final ($f$) magnonic state is given by
$I_{i\to f} ={2\pi}|\langle\psi_f|H_\mathrm{Int}|\psi_i\rangle|^2\delta(E_f-E_i)/\hbar$, where the interface exchange interaction between the metal and the AF is written as,
\begin{align*}
H_\mathrm{Int} = \sum_{\vec k\vec k'\vec q}
\left[ V^\alpha_{\vec q\vec k\vec k'}\alpha_{\vec q}c^\dag_{\downarrow\vec k} c_{\uparrow\vec k'}+ V^\beta_{\vec q\vec k\vec k'}\beta^\dag_{-\vec q}c^\dag_{\uparrow\vec k} c_{\downarrow\vec k'}\right] + h.c.,
\end{align*}

where $V^{\alpha,\beta}_{\vec q\vec k\vec k'}$ represents the electron-magnon scattering amplitudes for each magnon mode. For the current from the $\alpha$ magnons, we obtain -- under the assumption that the electrons are in thermal equilibrium and that only electrons with energies close to the Fermi energy $E_F$ can contribute significantly -- the general expression,
\begin{align}
I^\alpha &\nonumber= \frac{2\pi}{\hbar}\sum_{\vec k\vec k'\vec q}\int d\varepsilon\,
              (\varepsilon+\Delta\mu)
                \left[n_B\left(\frac{\varepsilon+\Delta\mu}{k_B T_e}\right) - n_I(\varepsilon)\right]\\
             & \times|V_{\vec q\vec k\vec k'}^\alpha|^2
             \delta(E_F{-}E_{\vec k})
             \delta(\varepsilon{-}\varepsilon_{\vec q}^\alpha)\delta(\varepsilon_{\vec q}^\alpha{+}E_{\vec k}{-}E_{\vec k'}),
\end{align}

where $n_I(\cdot)$ is the magnon distribution in the insulating layers.
At thermal equilibrium, $n_I(\varepsilon) = n_B(\varepsilon-\mu^\alpha)/k_BT_\alpha$.
Accordingly, we obtain an expression for the $\beta$ magnons. For the case where the ferromagnetically ordered planes of the NiO layer are in parallel to the interface, only one of the sublattices of NiO couples to the metal in our model and the matrix elements $V^{\alpha,\beta}_{\vec q\vec k\vec k'}$ are
\begin{align}
V^\alpha_{\vec q\vec k\vec k'} =\sqrt{2s}J_I\frac{\sin(k_x)\sin(k_x')}{N^{NM}_x}\frac{\eta(\vec q_\parallel)\delta_{\vec \kp+\vec q_\parallel,\vec k_\parallel'}u_{\vec q 0}}{\sqrt{N^I}},
\end{align}
where $\eta(\vec q_\parallel) = [e^{{-iq_z}/{\sqrt{6}}} + 2e^{{iq_z}/{2\sqrt{6}}}\cos({q_y}/{2\sqrt{2}})]$ and $u_{\vec q 0}$ is the amplitude of the magnonic eigenstate with momentum $\vec q$ at the interface. With this, we can write the equation for the $\alpha$ current slightly more convenient as
\begin{equation}
\begin{split}
    I^\alpha &=  \frac{s\pi J_I^2}{\hbar}
               \sum_{\Delta k_y\Delta k_z}\!\!\left\{
               |\eta(\Delta\vec k_\parallel)|^2 
               \left[\sum_{\overline{k_y} \overline{k_z}} \frac{\sin^2(k_x)}{\partial E_{\vec k}/\partial k_x}\frac{\sin^2(k_x')}{\partial E_{\vec k'}/\partial k_x'}\right]\right.\\
           & {\times}\left.  \sum_{q_x} |u_{\Delta \vec k_\parallel,q_x,0}|^2
             \! [\varepsilon^\alpha_{\vec q}+\Delta\mu]
              \!  \left[\!n_B\left(\frac{\varepsilon^\alpha_{\vec q}+\Delta\mu}{k_B T_e}\right) - n_I(\varepsilon^\alpha_{\vec q})\right]\!\right\}
\end{split}
\label{eq:Ialpha}
\end{equation}
where there is a dependence on the number of atomic layers of both, the AFI and the FI only in the second line of Eq. (\ref{eq:Ialpha}). Let us consider the situation $\Delta\mu=0$. Note that -- if we neglect the small effect of the NiO anisotropies $D_1$ and $D_2$ -- we have the following normalization condition
\begin{equation}
    \sum_{q_x} |u_{\Delta \vec k_\parallel,q_x,0}|^2 {h(q_x)}=(-1)^{N_M},
    \label{eq:normalization}
\end{equation}
where $h(q_x)=1$ for spin-($-1$) magnon operators ($\alpha$) and $h(q_x)=-1$ for spin-$1$ magnon operators($\beta$) mode, and $N_M$ is the number of atomic NiO planes. Furthermore, we consider now that magnons also to be in quasi equilibrium at temperature $T_M$, $n_I(\varepsilon^\alpha_{\vec q})=n_B({\varepsilon^\alpha_{\vec q}}/{k_B T_m})$.
{Now, we can compute the differential spin current, see Fig.~\ref{fig:diffcurrent} and the integrated spin current, see Fig.~\ref{fig:tempdep}.}
While $\varepsilon \left(n_B({\varepsilon}/{k_B T_e}) - n_B({\varepsilon}/{k_B T_m})\right)$ is not a constant as a function of $\varepsilon$, it gets smoother with increasing temperature. Consequently, the magnon spin transport at higher temperatures is mainly determined by the normalization condition Eq.~(\ref{eq:normalization}). 
Fig.~\ref{fig:tempdep} shows the amplitude of the spin current that increases with increasing temperature due to larger number of thermally excited magnons at higher temperature. Furthermore, we see that the differences in spin current for NiO layers of the same parity, tends to vanish with increasing temperature. This is due to Eq.~(\ref{eq:Ialpha}) being increasingly dominated by the influence amplitudes $u_{\Delta\kp,q_x,0}$ and those are subject to a normalization criterion as discussed above.

\textit{Discussion and summary}--
In this work, we considered perfect crystals of NiO which has higher N\'eel temperatures than polycrystalline NiO as experimentally investigated in \cite{Wangetal2014,Linetal2016}. Furthermore, the exchange coupling between the insulators in our model is strong which should result in a high blocking temperature so that the order of the NiO layer should be pinned by the magnetization of the FI. This is consistent with the result that the amplitude of the spin current is monotonously increasing with increasing temperature like in the experiments by Baldrati et al. \cite{Baldratietal2018} in contrast to the peak in the temperature dependence found by \cite{Linetal2016} for polycrystalline NiO. Our theoretical results heavily depend on the crystalline orientation in (111) direction. For other orientations, we expect a different behavior as the interface will most likely be compensated e.g. for NiO in (100) direction. Note that a dependence on the crystalline order was experimentally found for Cr$_2$O$_3$ layers \cite{Qinetal2020}.

Although the focus of this work was the antiferromagnet NiO, the results, especially the predicted even-odd effect are relevant for other materials. Namely, {enhanced} spin transport was already measured for CoO instead of NiO in \cite{Linetal2016}. The requirement for a precisely defined number of layers could also be achieved in magnetic van-der-Waals materials \cite{Sierraetal2021}. Here, specifically a layered antiferromagnet like CrI$_3$ \cite{Huangetal2017} is of interest.

We have computed the spin current between a ferromagnetic insulator and a heavy metal through a thin NiO layer in the clean limit and found a strong even-odd effect for the spin current. Moreover, at higher temperature, the influence of the NiO layer thickness apart from the even-odd effect vanishes. It remains an open question how details of a real system including damping will influence the spin current. However, our results suggest that as long as the magnetic structure in NiO is pinned by the FI, the even-odd effect could be observed in experiment with the remaining challenge that a sample needs to be prepared with a precise number of atomic planes.

\begin{acknowledgments}
This work was supported by the German Research Foundation (DFG) via the project no.~417034116 and by the Research Council of Norway through its Centres of Excellence funding scheme, Project No.~262633, ``Qu\-Spin''.
\end{acknowledgments}

\newcommand{\qp}{\vec q{_\parallel}}

\onecolumngrid
\section{Supplemental Material}
In this Supplemental Material, we show the details of the calculations of magnonic eigenstates.

\section{Holstein-Primakoff transformation}
We use a standard Holstein-Primakoff transformation to express the spin operators on the lattice sites by bosonic operators $a^{(\dag)}_{\vec r}$ for lattice sites $\vec r$ where in the presumed ground state the spins point in $z$ direction, i.e., on the A sublattice and $b^{(\dag)}_{\vec r}$  where it points in $-z$ direction (B planes). The Holstein-Primakoff transformation reads (we truncate after second order in the bosonic operators)
\begin{equation}
    S_r^z = \hbar(s-a^\dag_{\vec r}a_{\vec r}), S^+_{\vec r} = \hbar\sqrt{2}a_{\vec r}, S^-{\vec r} = \hbar\sqrt{2s}a^\dag_{\vec r}
\end{equation}
for the A sublattice and 
\begin{equation}
    S_r^z = \hbar(-s+b^\dag_{\vec r}b_{\vec r}), S^+_{\vec r} = \hbar\sqrt{2}b^\dag_{\vec r}, S^-{\vec r} = \hbar\sqrt{2s}b_{\vec r}
\end{equation}
for the B sublattice,
$S^x_{\vec r}=(S^+_{\vec r} + S^-_{\vec r})/2$,
$S^y_{\vec r}=(S^+_{\vec r} - S^-_{\vec r})/(2i)$.
\section{In-plane Fourier transform}
We perform a Fourier transform in the $yz$ plane and call the resulting operators $a^{(\dag)}_{j,\qp}$ and $b^{(\dag)}_{j,\qp}$, where $\qp$ is the in-plane momentum and $j$ the index of the plane in $x$ direction.

\subsection{Magnonic eigenstates of AFI-FI system}
We present here the formalism of the two-flavor Bogoliubov transformation for the case of $M$ atomic NiO layers
and $N$ atomic FI layers.
After the in-plane Fourier transform, the Hamiltonian can be denoted as
\begin{equation}
   H = \sum_{\qp} (\Psi_{\vec k_\parallel}^\dagger,\Psi^T_{-\vec k_\parallel}) \mathcal{H}_{\vec k_\parallel}\left(\begin{array}{c}\Psi_{\vec k_\parallel}\\ (\Psi_{-\vec k_\parallel}^T)^\dagger\end{array}\right)
   \label{eq:Hamil}
\end{equation}
where
$\Psi_\qp\!=\!(b^\dagger_{1,-\qp}, a_{2,\qp}, \ldots, b^\dagger_{M,-\qp}, a_{M{+}1, \qp},  a_{M{+}2,\qp},$
$ \ldots, a_{N{+}M,\qp})^T$
for an odd $M$ and
$\Psi_\qp\!=\!(a_{1,\vec k_\parallel},b^\dagger_{2,-\vec k_\parallel},\ldots,b^\dagger_{M,-\vec k_\parallel},a_{M{+}1,\vec k_\parallel},a_{M{+}2,\vec k_\parallel},\ldots,a_{N{+}M,\vec k_\parallel})^T$
for even $M$. The matrix Hamiltonian is thus,
\begin{equation}
    \mathcal{H}_{\vec k_\parallel} =\left(\begin{array}{cc}
    \mathcal{A}_\qp & \mathcal{B}\\
    \mathcal{B}     & \mathcal{A}_\qp \end{array}\right)
\end{equation}
where the matrix elements
\begin{equation}
    \mathcal{A}_{\vec k_\parallel}\!\! =\!\! \left(\!\begin{array}{ccccccccccc}
  d_1   & c_a    &        &       &             &        &     &      &       &       &\\
  c_a^* & d_a    & \ddots &       &             &        &     &      &       &       &\\
        & \ddots & \ddots & \ddots&             &        &     &      &       &       &\\
        &        & c_a^*  & d_a   & c_a         &        &     &      &       &       &\\ 
        &        &        & c_a^* & d_M         & c_{af} &     &      &       &       &\\
        &        &        &       & c_{af}^*    & d_{M+1}& c   &      &       &       &\\
        &        &        &       &             & c^*    & d   & c    &       &       & \\
        &        &        &       &             &        & c^* & d    & \ddots&       &  \\
        &        &        &       &             &        &     &\ddots& \ddots& \ddots&\\
        &        &        &       &             &        &     &      & c^*   & d     & c\\ 
        &        &        &       &             &        &     &      &       & c^*   & d_{N{+}M}
                                   \end{array}\!\!\right)
\end{equation}
and
\begin{equation}
    \mathcal{B} = \frac{S(D_1-D_2)}{2} \left(\begin{array}{cccc}
    1_{M\times M} & {0}_{M\times N} \\
    {0}_{N\times M}  & {0}_{N\times N}
     \end{array}\right),
\end{equation}
where $1_{M\times M}$ is the $M\times M$ identity matrix and ${0}_{i\times j}$ is a $i\times j$ matrix with zeros only. The coefficients in $\mathcal{A}_{\vec k_\parallel}$ are 
\begin{equation}
    \begin{split}
        d_1 & = S(J_1\gamma^{(1)}_\qp + 3J_2 - 3J_1),\\
        d_a & = S(J_1\gamma^{(1)}_\qp + 6J_2),\\
        c_a & = S(J_1\gamma^{(2)}_\qp+J_2\gamma_\qp^{(3)}),\\
        d_M & = S(J_1\gamma^{(1)}_\qp + 3J_2 - 3J_1) +3S_F(J_{I1}+J_{I2}),\\
        c_{af}     & = \sqrt{S_FS}(J_{I1}\gamma_\qp^{(2)}+J_{I2}\gamma_\qp^{(3)}),\\
        d_{M+1} & = S_FJ_F(\gamma^{(1)}_\qp-9) +3S(J_{I1}+J_{I2}),\\
        d & = S_FJ_F(\gamma_\qp^{(1)}-12) +2S_FK_F,\\
        c & =S_FJ_F\gamma^{(2)}_\qp,\\
        d_{M+N}& = S_FJ_F(\gamma_\qp^{(1)}-9) +2S_FK_F,
    \end{split}
\end{equation}
and
\begin{align}
\gamma^{(1)}_{\vec k_\parallel} & = 2\left[2\cos\left(\frac{k_y}{2\sqrt{2}}\right)
\cos\left(\frac{\sqrt{3}k_z}{2\sqrt{2}}\right)
+\cos\left(\frac{k_y}{\sqrt{2}}\right) \right],\\
\gamma^{(2)}_{\vec k_\parallel} & = e^{ik_z/\sqrt{6}}
+ 2e^{-ik_z/(2\sqrt{6})}\cos\left(\frac{k_y}{2\sqrt{2}}\right),\\
\gamma^{(3)}_{\vec k_\parallel} &  = \gamma^{(2)}_{-2\qp}.
\end{align}

Later, we determine the dynamical matrix by multiplying $\mathcal{H}_\qp$ with $\mathcal{G}$ where
\begin{equation}
    \mathcal{G}_{ij} = [\psi_{i,\qp},\psi^\dag_{j,\qp}]
\end{equation}
where $\psi^\dag_{i,\qp}$ and $\psi^\dag_{j,\qp}$ are the elements of
$(\Psi_\qp,(\Psi_\qp)^\dagger)$ and 
$((\Psi_\qp)^\dagger,\Psi_{-\vec k_\parallel})$,
respectively.
We diagonalize the matrices $\mathcal{G}\mathcal{H}_\qp$, i.e., we obtain the eigenenergies and amplitudes of the magnonic eigenstates on the lattice sites
by solving the eigenvalue problem \cite{Haakonetal}
\begin{equation}
    \mathcal{G}\mathcal{H}_\qp \vec u_\qp = \pm \varepsilon_\qp \vec u_\qp
    \label{eq:eigenvalueproblem}
\end{equation}
where $\varepsilon_\qp$ has to be one of the eigenenergies (for in plane magnon momentum $\qp$) and $\vec u_\qp$ is a vector of the amplitudes on each layer of the magnonic eigenstate.
The $\pm$ symbol is necessary due to the pseudo unitary behavior of the Bogoliubov transformation (while the magnon energies are all positive).
For the different solutions we formally assign an index $q_x$.
The creation and annihilation operators of the magnonic eigenstates will be denoted
$\alpha^{(\dag)}$ if they have more weight on the A layers and $\beta^{(\dag)}$ if they have more weight on the B layers).
We need the amplitude on the layer facing the metal, $u_{\qp,q_x,0}$ for the calculation of the spin current through the metal-insulator interface. We solve Eq.~(\ref{eq:eigenvalueproblem}) numerically for convenience.

\subsection{Normalization condition}
In general, each eigenstate of the dynamical matrix is a superposition of $a_{j,\qp}$, $a_{j,-\qp}^\dag$ ($j$ as index of A planes), $b_{j,\qp}$, and $b^\dag_{j,-\qp}$ (j as index of B planes).
However, the mixing of $a_{j,\qp}$ with $a_{j,-\qp}^\dag$ as well as $b_{j,\qp}$ with $b^\dag_{j,-\qp}$ is a consequence of the anisotropies of NiO. This turns out to be a negligible effect.
So we could instead consider $\mathcal{H}_\qp$ as two copies of a $(N+M)\times(N+M)$ matrix
yielding a more intuitive picture on how amplitudes on the sublattice are mixed in order to obtain magnonic eigenstates.
Then only $a_{j,\qp}$ and $b^\dag_{j,-\qp}$ are mixed ($\alpha$ or $\beta^\dag$) or $b_{j,\qp}$ and $a^\dag_{j,-\qp}$ ($\beta$ or $\alpha^\dag$)
Then for each magnon, identified by $q_x$, the normalization condition (following from the Bogoliubov transformation) reads
\begin{equation}
    \sum_j g(q_x,j) |u_{\Delta\qp,q_x,j}|^2 = h(q_x)
\end{equation}
where $j$ is now just the plane index in real space, $g(q_x,j)=1$ for $\alpha$ magnons (spin -1) on A layers and  $\beta$ magnons on B layers, whereas $g(q_x,j)=-1$ for $\alpha$ magnons on B layers and  $\beta$ magnons on A layers.
This normalization condition guarantees the correct bosonic commutation relations for the magnons.
If we sort the $\alpha^{(\dag)}$ and $\beta^{(\dag)}$ analogously to the $a^{(\dag)}$ and the $b^{(\dag)}$ operators being sorted in $(\Psi_{\vec k_\parallel}^T)^\dagger$ and $\Psi_{-\vec k_\parallel}$, we can write the transformation matrix $U$ with $\vec u_{\qp,q_x}$ as columns of $U$.
Then the normalization condition is $U^{\dag}\mathcal{G}U=\mathcal{G}$.
We find easily that we have similar normalization condition when summing over the magnon amplitudes at one layer,
\begin{equation}
    \sum_{q_x} g(q_x,0) |u_{\Delta\qp,q_x,0}|^2 = (-1)^{N_M}
\end{equation}
where we have plus on the right-hand side if layer 0 (the one facing the metal) is an A layer and minus if it is a B layer.

\end{document}